# Tunable Large Resonant Absorption in a Mid-IR Graphene Salisbury Screen

Min Seok Jang[1, †], Victor W. Brar[2,3, †], Michelle C. Sherrott[2], Josue J. Lopez[2], Laura K. Kim[2], Seyoon Kim[2], Mansoo Choi[1,4], and Harry A. Atwater[2,3]

† These authors contributed equally.

1) Global Frontier Center for Multiscale Energy Systems, Seoul National University, Seoul 151-744, Republic of Korea

2) Thomas J. Watson Laboratory of Applied Physics, California Institute of Technology, Pasadena, CA 91125

3) Kavli Nanoscience Institute, California Institute of Technology, Pasadena, CA91125

4) Division of WCU Multiscale Mechanical Design, School of Mechanical and Aerospace Engineering, Seoul National University, Seoul 151-742

**Enhancing the interaction strength between graphene and light is an important objective for those seeking to make graphene a relevant material for future optoelectronic applications. Plasmonic modes in graphene offer an additional pathway of directing optical energy into the graphene sheet, while at the same time displaying dramatically small optical confinement factors that make them an interesting means of coupling light to atomic or molecular emitters. Here we show that graphene plasmonic nanoresonators can**

**be placed a quarter wavelength from a reflecting surface and electronically tuned to mimic a surface with an impedance closely matched to freespace ($Z_0 = 377\ \Omega$). This geometry – known in early radar applications as a Salisbury screen – allows for an order of magnitude (from 2.3 to 24.5%) increase of the optical absorption in the graphene and provides an efficient means of coupling to the highly confined graphene plasmonic modes.**

The ability to interact strongly with light is important for a material to be useful in optics-based applications. Monolayer graphene exhibits a number of interesting optical phenomena including a novel photo-thermoelectric effect,[4,5] strong non-linear behavior,[6,7] and the potential for ultra-fast photodetection.[8] However, the absolute magnitude of these effects is limited by the amount of light absorbed by the graphene sheet, which is typically 2.3% at infrared and optical frequencies[9,10] - a small value that reflects the single atom thickness of graphene. To increase total overall graphene-light interaction, a number of novel light scattering and absorption geometries have recently been developed. These include coupling graphene to resonant metal structures[11-16] or optical cavities where the electromagnetic fields are enhanced[17-19], or draping graphene over optical waveguides to effectively increase the overall optical path length along the graphene.[20,21] While those methods rely on enhancing interband absorption processes, graphene can also be patterned and doped so as to excite plasmonic modes that display strong resonant absorption in the terahertz to mid-infrared regime.[22-26] The plasmonic modes are highly sensitive to their environment, and they have been shown to display large absorption when embedded in liquid salts[22,27] or by sandwiching dopants between several graphene layers.[26] However without blocking the transmission of light, it is not possible to achieve unity absorption in these previously demonstrated geometries[22,27][26]. Moreover, in order to access nonlinear or high frequency modulation as well as the high confinement factors characteristic of graphene plasmons, device geometries with open access to the graphene surface that operate with field effect gating at low doping are desirable.

Plasmonically active metallic and semiconductor structures can achieve near-perfect absorption of radiation at specified frequencies using a resonant interference absorption method.[28-32] The electromagnetic design of these structures derives in part from the original Salisbury screen design, but with the original resistive sheet replaced by an array of resonant metal structures used to achieve a low surface impedance at optical frequencies. The high optical interaction strength of these structures has made them useful in such applications as chemical sensing,[29,33] and it was recently proposed that similar devices could be possible using graphene to achieve near perfect absorption from THz to Mid-IR.[34,35] Such a device would offer an efficient manner of coupling micron-scale freespace light into nanoscale plasmonic modes, and would further allow for electronic control of that in-coupling process. In this Letter, we construct a device based on that principle, using tunable graphene nanoresonators placed a fixed distance away from a metallic reflector to drive a dramatic increase in optical absorption into the graphene.

A schematic of our device is shown in Figure 1a. A graphene sheet grown using chemical vapor deposition on copper foil is placed on a 1μm thick low stress silicon nitride ($SiN_x$) membrane with 200nm of Au deposited on the opposite side that is used as both a reflector and a backgate electrode. Nanoresonators with widths ranging from 20-60nm are then patterned over 70×70 μm$^2$ areas into the graphene using 100keV electron beam lithography (see Methods). An atomic force microscope (AFM) image of the resulting graphene nanoresonators is shown in the inset of Fig. 1b. The device was placed under a Fourier transform infrared (FTIR) microscope operating in reflection mode, with the incoming light polarized perpendicular to the resonators. The carrier density of the graphene sheet was varied *in situ* by applying a voltage across the SiN$_x$ between the gold and the graphene, and the resulting changes in resistance were continuously monitored using source and drain electrodes connected to the graphene sheet (Fig 1b). The carrier density of the graphene nanoresonators was determined from experimentally measured resonant peak frequencies (see Section I & II in Supplementary Information).

The total absorption in the device – which includes absorption in the SiN$_x$ and the graphene resonators - is determined from the difference in the reflected light from the nanoresonator arrays and an adjacent gold mirror. For undoped and highly doped 40nm nanoresonators, the total absorption is shown in Figure 2a, revealing large absorption at frequencies below 1200cm$^{-1}$, as well as an absorption peak that varies strongly with doping at 1400cm$^{-1}$ and a peak near 3500cm$^{-1}$ that varies weakly with doping. In order to distill absorption features in the graphene from the environment (i.e., SiN$_x$ and Au back reflector), we plot the difference in absorption between the undoped and doped nanoresonators, as shown in Figure 2b for 40nm nanoresonators. This normalization removes the low frequency feature below 1200cm$^{-1}$, which is due to the broad optical phonon absorption in the SiN$_x$ and is independent of graphene doping. The absorption feature at 1400cm$^{-1}$, however, shows a dramatic dependence on the graphene sheet carrier density, with absorption into the graphene nanoresonators varying from near 0% to 24.5% as the carrier density is raised to $1.42 \times 10^{13}$/cm$^2$. Because the absorption increases with carrier density, we associate it with resonant absorption in the confined plasmons of the nanoresonators.[22-25,36] In Figure 2b we also see that absorption at 3500cm$^{-1}$ exhibits an opposite trend relative to the lower energy peak, with graphene-related absorption decreasing with higher carrier density. This higher energy feature is due to interband graphene absorption, where electronic transitions are Pauli blocked by state filling at higher carrier densities.[37] For spectra taken from the bare, gate-tunable graphene surface, this effect leads to ~8% absorption, roughly twice the intensity observed from patterned areas. Finally, in Figure 2c, we investigated the graphene nanoresonator absorption as the resonator width is varied from 20 to 60nm at fixed carrier density. This figure shows that the lower energy, plasmonic absorption peak has a strong frequency and intensity dependence on resonator width, with the maximum absorption occurring in the 40nm ribbons.

The carrier density dependent plasmonic dispersion of this system is shown in Figure 3a. The observed resonance frequency varies from 1150-1800cm$^{-1}$, monotonically increasing with larger carrier densities and smaller resonator widths. The plasmon energy asymptotically approaches ~1050cm$^{-1}$ due to a polar phonon in the SiN$_x$ that strongly reduces the dielectric function of the substrate at that energy.[38] This coupling between the substrate polar phonon and the graphene plasmon has also been previously observed in back-gated SiO$_2$ devices.[23,25,39] In Figure 3b we plot the intensity of the plasmonic absorption as a function of frequency at varying carrier densities, revealing that for all carrier densities, the maximum in absorption always occurs at 1400cm$^{-1}$. ,

The experimental behavior observed in Figures 2 and 3 has some similarities with graphene plasmonic resonators patterned on back-gated SiO$_2$ devices, however there are some significant differences. Most notably, the absolute absorption observed in this device is one order of magnitude greater that what has previously been observed. Furthermore the maximum absorption in this device always occurs near 1400cm$^{-1}$, in contrast to previous graphene plasmonic devices where lower frequency resonances showed greater intensity due to fewer loss pathways and better *k*-vector matching between the graphene plasmons and freespace light.[23,25]

These new absorption features can be understood by considering the role of the gold reflector. At 1400cm$^{-1}$ the optical path length of the SiN$_x$ is $\lambda/4n$ and the gold reflector creates a standing wave between the incident and reflected light that maximizes the electric field on the SiN$_x$ surface. As a consequence, when the graphene nanoresonators are tuned to absorb at 1400cm$^{-1}$, a double resonance condition is met, and the dissipation of the incoming radiation is greatly enhanced. In order to illustrate the role of the interference effect, the frequency dependence of the electric field intensity on the bare nitride surface is plotted as a dashed curve in Figure 2c. As can be seen in this figure, the intensity of the plasmonic absorption displays a frequency dependence that is similar to the calculated field intensity.

Full wave finite element electromagnetic simulations are performed in order to better understand the performance of our device and the underlying mechanisms driving the large observed absorption.[23] The conductivity of the graphene sheet is modeled using the local random phase approximation[40] with the intraband scattering rate $\Gamma$ including both scattering by impurities $\Gamma_{imp}$ and by optical phonons $\Gamma_{oph}$. By analyzing the absorption peak width when the resonance energy is much lower than the graphene optical phonon energy (~1600cm$^{-1}$), the impurity scattering rate is approximated to be $\Gamma_{imp}=ev_F/\mu\sqrt{n\pi}$, with a carrier mobility of 550cm$^2$/Vs.[36] The rate of optical phonon scattering is estimate from the theoretically obtained self-energy $\Sigma_{oph}(\omega)$, as $\Gamma_{oph}(\omega)=2\text{Im}[\Sigma_{oph}(\omega)]$.[25,36,41] In order to match the calculations to our experimentally determined spectra, we multiply the theoretical spectrum by a constant factor of 0.72. This factor accounts for experimental imperfections in the device such as electronically isolated resonators caused by cracks in the graphene sheet, or resonators that contain a graphene grain boundary. Our resulting theoretical curves for the frequency and intensity dependence of the resonant absorption are shown in Figs. 3a and 3b, respectively. As seen in Figure 3b, the theory and the measurement show similar features - a maximum plasmonic absorption consistently occurs around 1400cm$^{-1}$ for a given charge density regardless of the resonator width. The field profiles from our calculations are shown in Fig. 3c, revealing a strong plasmonic response in the graphene nanoresonators for the λ/4$n$ condition where the electric field is maximized on the surface and the resonators match the correct resonance conditions.

A more complete understanding of the large resonant absorption observed in this graphene Salisbury screen comes from viewing the effect in terms of impedance matching, where the graphene metasurface is modified in such a way that it mimics a load whose admittance is close to the free space wave admittance $Y_0 = \sqrt{\epsilon_0/\mu_0}$, and thus allows for all incident light to be absorbed in the graphene sheet.[1,3] This description is diagramed in the inset of Figure 1a. To understand this model, we can consider the effective admittance of a thin layer of thickness $\tau$ and admittance $Y_{GR} = \sqrt{\epsilon_{GR}/\mu_{GR}}$ sitting atop a dielectric with thickness $d$ and admittance $Y_{SiN_x}$ deposited on a reflecting mirror. For frequencies such that $d = m\lambda/4$ and for $\tau \ll 1$, the total effective admittance of the stack is given by $Y = -i\omega\epsilon_{GR}\tau$ (see Section IV in Supplementary Information). For normally incident light, the amount of absorption is given by $A = 1 - |(Y_0 - Y)/(Y_0 + Y)|^2$ when the layer is located a quarter wavelength away from the back reflector.[3] Thus, the absorption approaches unity as the relative admittance $Y/Y_0$ approaches 1.

Typically, the admittance of an unpatterned graphene sheet is quite low, and equivalent to its sheet conductivity $\sigma$. Thus for unpatterned graphene, $Y = \sigma \approx e^2/4\hbar = \pi\alpha Y_0 \approx 0.023 Y_0$ when the photon energy is sufficiently higher than the Pauli-blocked interband transition energies, where $\alpha$ is the fine structure constant. As a result, the absorption by a pristine graphene monolayer in the Salisbury screen configuration can be calculated as $A \approx 8.8\% \approx 4\pi\alpha$ which is consistent with our experimental observations of the higher energy feature at 3500cm$^{-1}$ shown in Figure 2b.

With optical resonators patterned into the graphene layer, however, the surface electric admittance can be dramatically increased. When the resonators are sparsely spaced so that they barely interact with each other, one can obtain the effective permittivity of the resonator array by simply multiplying the spatial density of the resonators by the polarizability of an individual resonator $a(\omega)$. The admittance is then $Y = -i\omega a(\omega)/S$, where $S$ is the area of the unit cell. On resonance, there is a dramatic increase in Im[$a$], while Re[$a$] crosses zero.[35] Recognizing that the absorption cross-section of a dipole is $\sigma_{Abs} = (\omega/c)\text{Im}[a/\varepsilon_0]$, the surface admittance is given by $Y = (\sigma_{Abs}/S)Y_0$ on resonance. This is physically intuitive because it says that complete absorption occurs when the absorption cross section of the resonator array is large enough to cover the entire surface. As the resonators become closer to each other, the resonance frequency redshifts due to inter-resonator coupling, yet the condition for perfect absorption remains valid.[35] For our device at its highest doping level, $\sigma_{Abs}/S$ is estimated to be $0.13Y_0$, which is much higher than $\pi\alpha$, and this allows for the large absorption we observe in our graphene nanoresonators shown in Figure 2. Increasing carrier density leads to better coupling between the incoming light and the graphene plasmons, resulting in a stronger plasmon resonance. Therefore, at a given resonance frequency, higher doping enhances the absorption performance as seen in Figure 3b and S6.

Finally, we point out that the resonant absorption can be further increased if the resistive damping in the graphene is reduced. In Figure 4a, we plot the calculated carrier mobility dependence of the surface admittance for an array of graphene nanoribbons on a 1μm SiN$_x$/Au layer. The highest achieved carrier density $1.42\times10^{13}$/cm$^2$ is assumed, and the width of the ribbons is chosen to be 40nm in order to match the plasmon resonance with the quarter wavelength condition of the $SiN_x$ layer (~1400cm$^{-1}$). Because the resonator absorption cross-section increases as the graphene becomes less lossy, the resonant surface admittance increases with increasing mobility and crosses the free space admittance $Y_0$ at a carrier mobility of $\mu \approx$ 4,000cm$^2$/Vs. As $Y$ exceeds $Y_0$, the maximum absorption starts decreasing. However, it should be noted that in this high mobility regime, perfect absorption can still be achieved by shifting the quarter wavelength condition from the plasmon resonance frequency via changing the $SiN_x$ thickness in order to decrease the coupling between the free wave and the graphene plasmon. To illustrate this, Figure 4b shows the simulated peak absorption in the same resonator array as a function of both the mobility and the thickness of the nitride layer. Indeed, for $Y > Y_0$ the perfect absorption occurs at two different thickness values: one thinner and another thicker than 1 μm. This deviation becomes larger as the graphene mobility increases, and for mobilities reaching 10,000cm$^2$/Vs the device will show total absorption for nitride layers with thicknesses of 700nm or 1.3 μm.

In summary, we have experimentally demonstrated that graphene plasmonic resonators placed a quarter wavelength away from a back reflector can absorb almost 25% of incoming Mid infrared light - more than 10 times higher than the case of unpatterend graphene without a reflector (~2.3%). The frequency and the amount of absorption can be largely tuned by controlling the plasmon resonance of the nanoresonators via electrostatic gating or varying the resonator size.  This strong optical response allows for graphene to be considered relevant as a serious material to be used in optoelectronic devices.  Furthermore, our modeling predicts that modestly increasing the graphene mobility or decreasing the resonator line roughness can lead to 100% absorption, a tangible and important goal.  Finally, these results clearly demonstrate that the extremely small mode volumes of graphene plasmonic modes can be made accessible to free space probes despite the large discrepancies in wavelength that suppress such coupling. Because the large light confinement of graphene plasmons allow them to couple efficiently to nearby dipole emitters, this technique could indirectly allow for a robust interaction between molecular scale emitters and free space light.

## Methods

**Device Fabrication.** $SiN_x$ membranes were obtained commercially from Norcada, part #NX10500F. Electron beam lithography at 100keV is used to pattern nanoresonator arrays in PMMA spun coated onto the devices, and the pattern is transferred to the graphene via an oxygen plasma etch. Our resonators have widths varying from 20 – 60nm, with 9:1 aspect ratios and a pitch of 2-2.5 times the width. The resonators are spanned perpendicularly by graphene crossbars of a width equal to the nanoresonator width. This aids conductivity across the patterned arrays despite occasional cracks and domain boundaries in the CVD graphene sheet.

## Figure Captions

**Figure 1. Schematic of experimental device.** (a) $70 \times 70 \mu m^2$ graphene nanoresonator array is patterned on 1μm thick silicon nitride ($SiN_x$) membrane via electron beam lithography. On the opposite side, 200nm of gold layer is deposited that serves as both a mirror and a backgate electrode. A gate bias was applied across the $SiN_x$ layer in order to modulate the carrier concentration in graphene. The reflection spectrum was taken using a Fourier Spectrum Infrared (FTIR) Spectrometer attached to an infrared microscope with a 15X objective. The incident light

was polarized perpendicular to the resonators. The inset schematically illustrates the device with the optical waves at the resonance condition. (b) DC resistance of graphene sheet as a function of the gate voltage. The inset is an atomic force microscope image of 40 nm nanoresonators.

**Figure 2. Gate-induced modulation of absorption in graphene nanoresonator arrays.** (a) The total absorption in the device for undoped (red dashed) and highly hole doped (blue solid) 40nm nanoresonators. Absorption peaks at 1400cm$^{-1}$ and a peak at 3500cm$^{-1}$ are strongly modulated by varying the doping level, indicating these features are originated from graphene. On the other hand, absorption below 1200cm$^{-1}$ is solely due to optical phonon loss in SiN$_x$ layer. (b) The change in absorption with respect to the absorption at the charge neutral point (CNP) in 40nm wide graphene nanoresonators at various doping levels. The solid black curve represents the absorption difference spectrum of bare (unpatterned) graphene. (c) Width dependence of the absorption difference with the carrier concentration of 1.42×10$^{13}$cm$^{-2}$. The width of the resonators varies from 20 to 60nm. The dashed curve shows the theoretical intensity of the surface parallel electric field at SiN$_x$ surface when graphene is absent. Numerical aperture of the 15X objective (0.58) is considered.

**Figure 3. Dispersion and peak absorption of plasmon resonances in graphene nanoresonator arrays.** (a) Peak frequency as a function of resonator width. Solid curves and the symbols plot the theoretical and measured peak frequencies respectively. The width of the nanoresonators are determined by AFM measurements. (b) Frequency dependence of the maximum absorption difference with varying doping level (symbols). The solid curves indicate the theoretical values obtained from finite element electromagnetic simulations multiplied by a constant factor 0.72 which takes into account the fabricational imperfections such as dead resonators. (c) Electric field profile of a 40nm graphene nanoresonator with the highest achieved carrier density ($1.42 \times 10^{13} cm^{-2}$), obtained from an electromagnetic simulation assuming normal incidence. The quarter wavelength condition and plasmon resonance coincide at $1400 cm^{-1}$.

**Figure 4. Carrier mobility dependence**. (a) Dependence of normalized surface admittance $Y/Y_0$ of 40nm graphene nanoribbon array on resonance (red) and the maximum absorption (right) on the carrier mobility $\mu$ (intraband scattering rate $\Gamma = ev_F/\mu\sqrt{n\pi}$ ). The thickness of the $SiN_x$ layer and the pitch are assumed to be 1um and 40nm, respectively. The admittance is monotonically increases as the mobility increases, but is not directly proportional to it due to the loss in $SiN_x$ substrate. 100% absorption occurs at $\mu \approx 4,000 cm^2/Vs$. (b) Maximum absorption in the device as a function of the $SiN_x$ thickness and the mobility. Impedance matching condition ($Y = Y_0$) is indicated as the grey dashed line. The red dotted curve indicates the condition for perfect absorption.

## Acknowledgements

We gratefully acknowledge support from the Air Force Office of Scientific Research Quantum Metaphotonic MURI program under grant FA9550-12-1-0488 and use of facilities of the DOE "Light-Material Interactions in Energy Conversion" Energy Frontier Research Center (DE-SC0001293).  M. S. Jang and M. Choi acknowledge support from the Global Frontier R&D Program on Center for Multiscale Energy Systems funded by the National Research Foundation under the Ministry of Science, ITC & Future Planning, Korea (2011-0031561, 2011-0031577). M.S. Jang acknowledges a post-doctoral fellowship from the POSCO TJ Park Foundation. V.W. Brar gratefully acknowledges a post-doctoral fellowship from the Kavli Nanoscience Institute. M.C. Sherrott gratefully acknowledges graduate fellowship support from the Resnick Sustainability Institute at Caltech. S. Kim and M.S. Jang acknowledges support from a Samsung Fellowship.

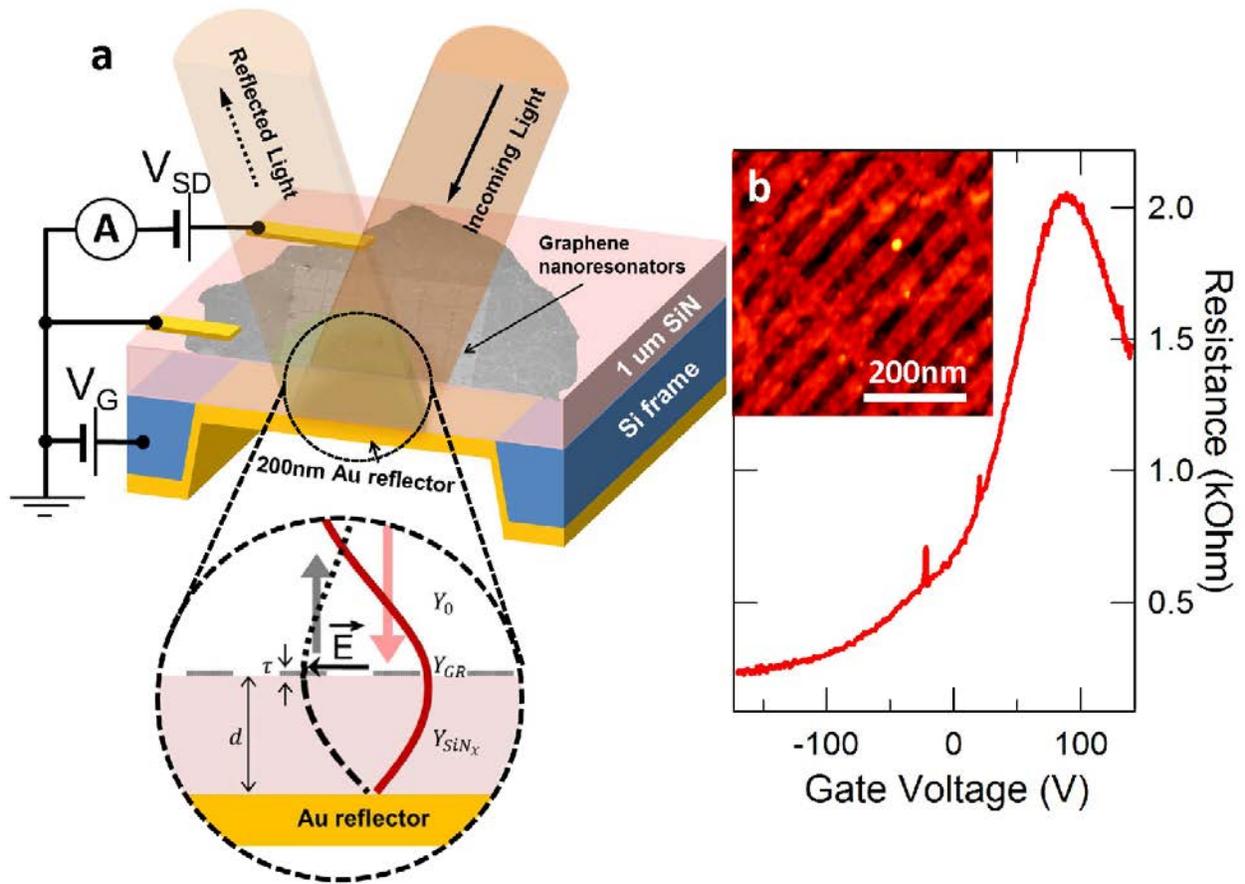

**Figure 1. Schematic of experimental device.**

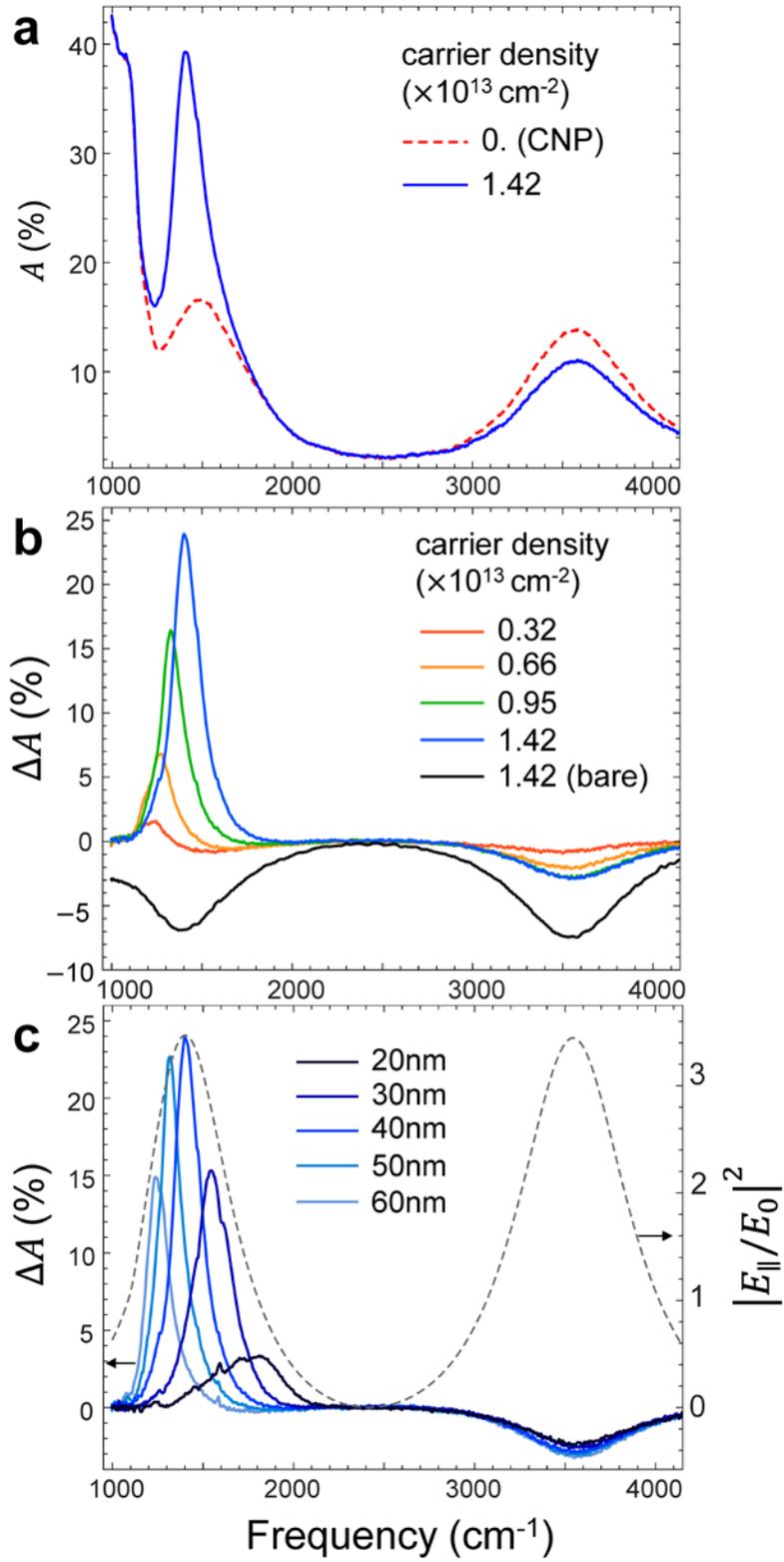

**Figure 2. Gate-induced modulation of absorption in graphene nanoresonator arrays.**

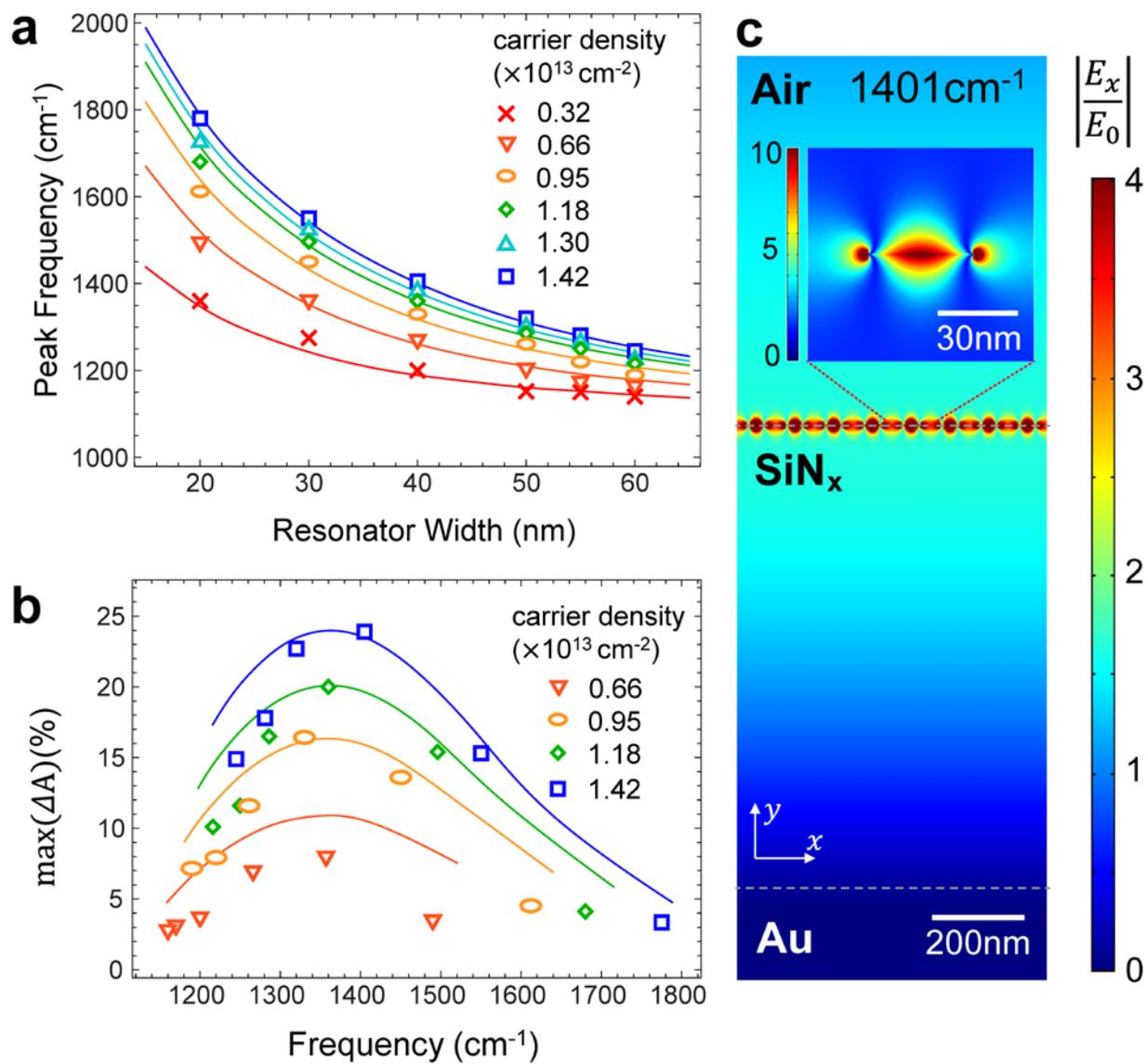

**Figure 3. Dispersion and peak absorption of plasmon resonances in graphene nanoresonator arrays.**

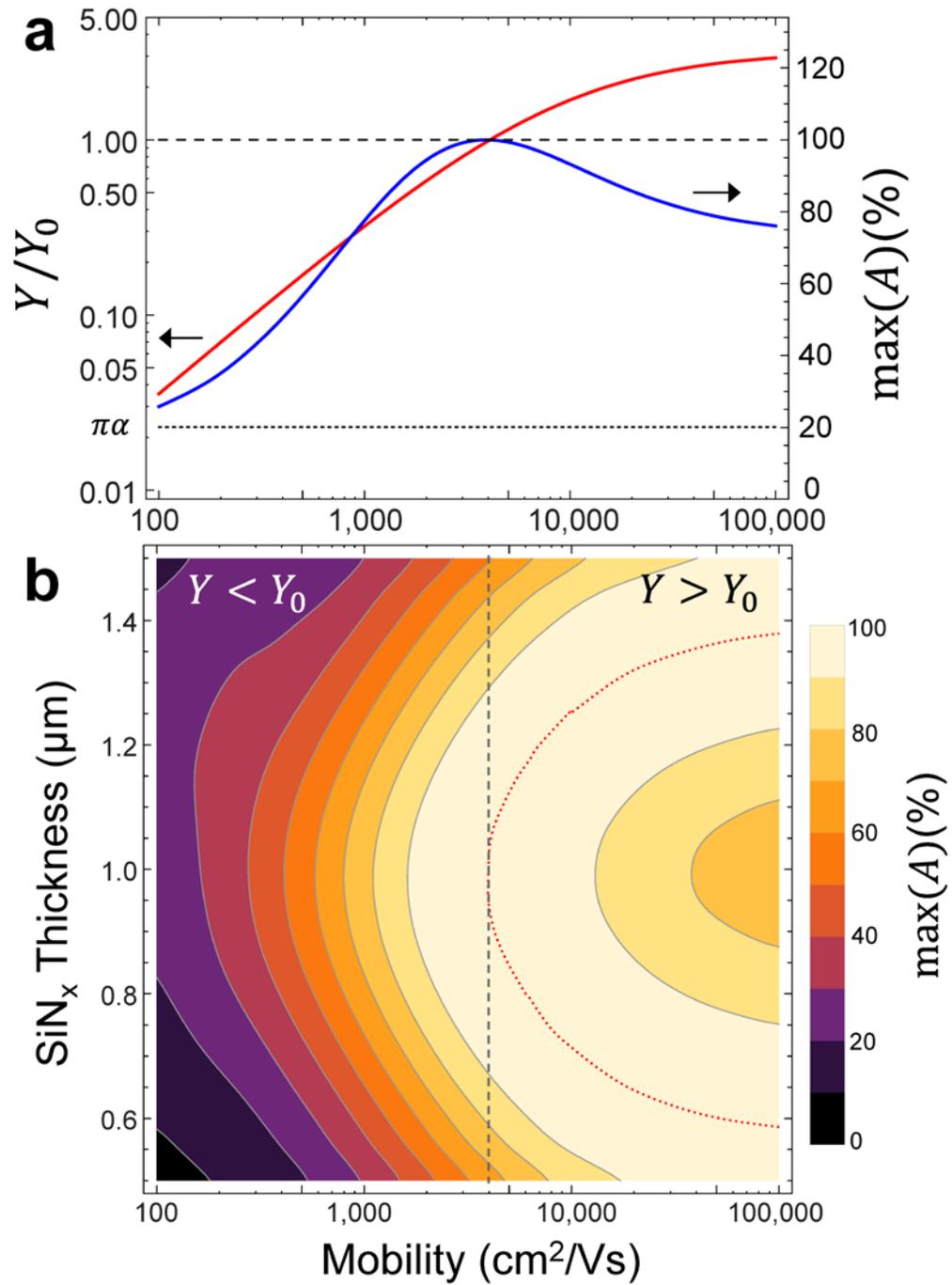

**Figure 4. Carrier mobility dependence**.

# Supplementary Information:

# Tunable Large Resonant Absorption in a Mid-IR Graphene Salisbury Screen

Min Seok Jang[1, †], Victor W. Brar[2,3, †], Michelle C. Sherrott[2], Josue J. Lopez[2], Laura K. Kim[2], Seyoon Kim[2], Mansoo Choi[1,4], and Harry A. Atwater[2,3]

† These authors contributed equally.

1) Global Frontier Center for Multiscale Energy Systems, Seoul National University, Seoul 151-747, Republic of Korea

2) Thomas J. Watson Laboratory of Applied Physics, California Institute of Technology, Pasadena, CA 91125

3) Kavli Nanoscience Institute, California Institute of Technology, Pasadena, CA91125

4) Division of WCU Multiscale Mechanical Design, School of Mechanical and Aerospace Engineering, Seoul National University, Seoul 151-742

# I. Electromagnetic Simulations

We solve Maxwell's equation by using finite element method. Graphene is modeled as a thin layer of the thickness $\tau$ and impose the relative permittivity $\epsilon_G = 1 + i\sigma/(\epsilon_0 \omega \tau)$. In actual calculation, $\tau$ is chosen to be 0.1 nm which shows good convergence with respect to the $\tau \to 0$ limit. The complex optical conductivity of graphene $\sigma(\omega)$ is evaluated within local random phase approximation.[1]

$$\sigma(\omega) = \frac{2ie^2 T}{\pi\hbar(\omega + i\Gamma)} \log\left[2\cosh\left(\frac{E_F}{2T}\right)\right] + \frac{e^2}{4\hbar}\left[H\left(\frac{\omega}{2}\right) + \frac{4i\omega}{\pi}\int_0^\infty d\eta \frac{H(\eta) - H\left(\frac{\omega}{2}\right)}{\omega^2 - 4\eta^2}\right],$$

where

$$H(\eta) = \frac{\sinh(\eta/T)}{\cosh(E_F/T) + \cosh(\eta/T)}.$$

Here, the temperature $T$ is set as 300K. The intraband scattering rate $\Gamma$ takes into account scattering by impurities $\Gamma_{imp}$ and by optical phonons $\Gamma_{oph}$. By analyzing the absorption peak width when the resonance energy is much lower than the graphene optical phonon energy (~1600cm$^{-1}$), the impurity scattering rate can be approximated to be $\Gamma_{imp} = ev_F/\mu\sqrt{n\pi}$ with the mobility $\mu = 550$cm$^2$/Vs.[2] The rate of optical phonon scattering is estimated from theoretically obtained self-energy $\Sigma_{oph}(\omega)$, as $\Gamma_{oph}(\omega) = 2\text{Im}[\Sigma_{oph}(\omega)]$.[2-4] The frequency dependent dielectric functions of Au and SiN$_x$ are taken from Palik [5] and Cataldo et al. [6], respectively. Finally, a constant factor, which accounts for experimental imperfections such as dead resonators in the actual device, is multiplied to the simulated spectra. This degradation factor is determined to be 0.72 by comparing simulation and measurement. Figure S1 shows that the resulting theoretical absorption spectra reproduce quite well the experimental data.

## II. Determination of Carrier Density

The carrier density of the nanoresonators was determined by fitting the peak frequencies of the simulated absorption spectra to the experimentally measured absorption peaks of resonators fabricated with different widths. The resulting carrier density values are comparable to those calculated using a simple parallel plate capacitor model with a 1 μm thick $SiN_x$ dielectric, as shown in Figure S2a, yet there is some deviation. We can attribute the discrepancies to a number of possible effects. First, our $SiN_x$ membranes were obtained from a commercial supplier (Norcada) and their stoichiometry and resulting DC dielectric constant, ☐, is not precisely known. This allows for a range of possible values for ☐, which can lead to significant differences in the induced carrier density in a graphene device. Second, our measurements were performed under FTIR purge gas (free of $H_2O$ and $CO_2$), but atmospheric impurities were likely still present during the measurements. Those types of impurities have previously been shown to induce hysteresis effects in the conductance curves of graphene FET devices,[7-10] and we observe similar behavior in our devices, as shown in Figure S2b. Because the concentration of those impurities can depend on the applied gate bias, they can also alter the carrier density vs. gate bias curves, and in Figure S2a we have included a theoretical estimate of those effects.[9] In addition to atmospheric impurities, the $SiN_x$ surface itself can contain charge traps that fill or empty with the applied gate bias. Such charge traps could induce anomalous behavior in the conductance curves of the graphene FET devices, similar to what has been observed in the presence of metallic impurities.[11] Finally, we note that we have removed some of the graphene surface area in the process of fabricating the nanoresonators. This difference in total available surface area should alter the carrier density dependence assumed by the capacitor model, such that more charge is likely packed into a smaller area. In Figure S2a we have provided a simple estimate of this effect based on the assumption that an equal amount of induced carriers are distributed equally across the smaller available surface area, leading to larger carrier densities. However, theoretical predictions have shown that the extra carrier density should preferably accumulate on the edges of the graphene nanoresonators, and thus alter their plasmonic resonances in more sophisticated ways.[12]

## III. Peak Width Analysis

In Figure S3a, we plot the full with at half maximum (FWHM) of the absorption peaks of graphene nanoresonator arrays with various sizes and doping levels. The linewidth, which can be interpreted as the plasmon scattering rate, almost monotonically increases with increasing resonance frequency and decreasing resonator width. The lifetime of plasmon is estimated as 10-50 fs from inverse linewidth.

When the substrate medium is lossless and dispersionless, the scattering rate of graphene plasmon is simply equal to the electron scattering rate. However, in our sample, the interaction with $SiN_x$ substrate polar phonons results in a deviation of the plasmon scattering rate from the electron scattering rate. Therefore, we extract the intraband electron scattering rate ($\Gamma$) by fitting the FWHM of the simulated spectrum to the measured plasmon linewidth, as shown in figure S3b.

We found that there is no noticeable difference in the electron scattering rates among nanoresoantors wider than 40nm. Because those nanoresonators oscillate at frequencies much lower than graphene optical phonon (~1600cm$^{-1}$), the dominant damping mechanism in this regime is scattering from impurities.[2,3] The average carrier mobility $\mu$, converted from the electron scattering rate via $\mu = ev_F/\Gamma\sqrt{n\pi}$, is determined as 550cm$^2$/Vs with standard deviation 50cm$^2$/Vs. On the other hand, at frequencies higher than 1600cm$^{-1}$, the electron scattering rate of 20nm nanoresonators dramatically increases as the carrier density increases (and thus the plasmon frequency increases), possibly due to coupling with graphene optical phonons.

## IV. Derivation of Surface Admittance of a Thin Layer

Consider a thin layer of thickness $\tau$ and admittance $Y_{GR}$ sitting atop a dielectric with thickness $d$ and admittance $Y_{SiN_x}$ deposited on a reflecting mirror as diagramed in the inset of Figure 1a. For normally incident light, the effective surface admittance of the stack is given by [13,14]

$$Y = Y_{GR} \frac{Y'_{SiN_x} - iY_{GR}\tan(k_1\tau)}{Y_{GR} - iY'_{SiN_x}\tan(k_1\tau)},$$

where $Y_{GR} = \sqrt{\epsilon_{GR}/\mu_{GR}}$ and $k_1 = \omega\sqrt{\epsilon_{GR}\mu_{GR}}$ are the wave admittance and the wavevector inside the thin sheet, respectively. $Y'_{SiN_x}$ is the effective admittance of the dielectric as viewed from the position of the sheet, and is given by $Y'_{SiN_x} = Y_{SiN_x}\cot(k_2 d)$, where $k_2$ is the wavevector inside the SiN$_x$ layer. For frequencies such that $d = m\lambda/4$ and for $k_1\tau \ll 1$, then $Y'_{SiN_x} \ll Y_{GR}$ and $\tan(k_1\tau) \to k_1\tau$, and the above equation reduces to $Y = -i\omega\varepsilon\tau$.

## V. Calculation of Surface Admittance of Graphene Nanoresonator Arrays

The surface admittance $Y = -i\omega\varepsilon\tau$ of a graphene nanoresonator array is equivalent to its effective sheet conductivity $\sigma_{eff}$, which can be evaluated from the far-field transmission and reflection coefficients. Consider a homogeneous thin film of conductivity $\sigma_{eff}$ placed on the interface ($z = 0$) between air ($z < 0$) and SiN$_x$ ($z > 0$) and a plane wave polarized along $x$ direction is normally incident on the surface. The surface parallel electric field is continuous $E_x^{0+} = E_x^{0-}$ at the interface, while the magnetic fields are discontinuous due to the surface current, $H_y^{0+} - H_y^{0-} = \sigma_{eff}E_x(z=0) = YE_x(z=0)$. From these boundary conditions, the transmission ($t$) and reflection ($r$) coefficients satisfy the following equations,

$$1 + r = t,$$

$$(1 - r) - n_{SiN_x}t = \left(\frac{Y}{Y_0}\right)t,$$

where $n_{SiN_x}$ is the refractive index of SiN$_x$. The normalized surface admittance $Y/Y_0$ is then solely written in terms of transmission coefficient,

$$\frac{Y}{Y_0} = \frac{2}{t} - 1 - n_{SiN_x}.$$

Because the fields of graphene plasmons are tightly confined near the surface with characteristic decay length similar to the width of the nanoresonators, we record the electric field of the transmitted wave at a position sufficiently far from the surface ($z_0 = 1$um) in order to exclude the evanescent field of graphene plasmons. The far field transmission coefficient is then obtained by accounting for the propagation factor

$$t = \frac{E_x(z_0)\exp[-in_{SiN_x}k_0 z_0]}{E_0(0)},$$

where $E_0$ is the electric field of incident wave at the surface.

Figure S4 plots the resulting complex surface admittance of a graphene nanoribbon array as a function of frequency. As in figure 4, both ribbon width and the spacing between the ribbons are set as 40nm. On resonance, Im[$Y$] crosses zero, while Re[$Y$], which is directly proportional to the absorption cross section $\sigma_{Abs}$ of individual resonator, has its maximum. As graphene becomes less lossy, the plasmon resonance gets sharper and stronger.

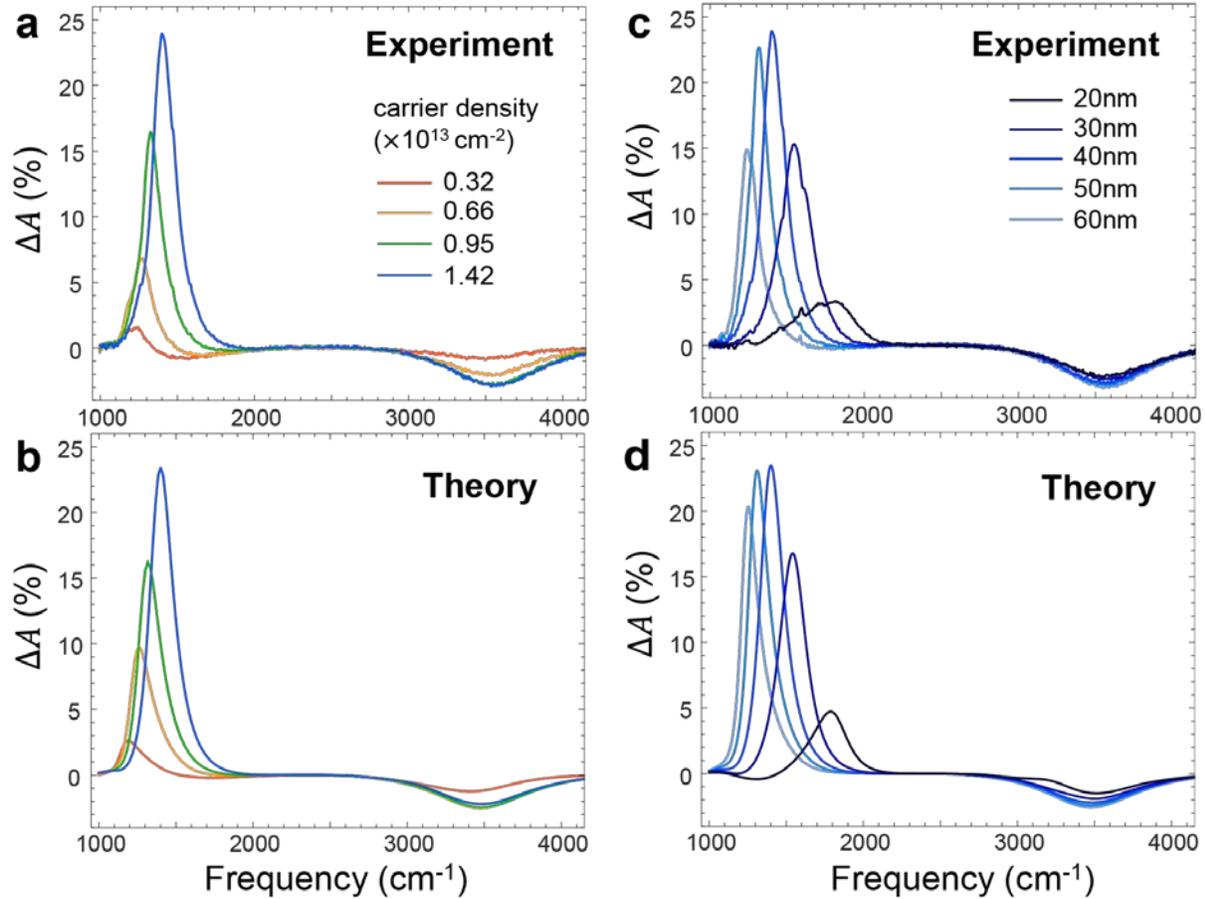

Figure S1: **Comparison between experimental and theoretical absorption spectra.** (a) Experimental and (b) theoretical change in absorption with respect to the absorption at the charge neutral point (CNP) in 40nm wide graphene nanoresonators at various doping levels. (c) Experimental and (d) theoretical absorption difference spectra with the carrier concentration of $1.42 \times 10^{13}$ cm$^{-2}$. The width of the resonators varies from 20 to 60nm.

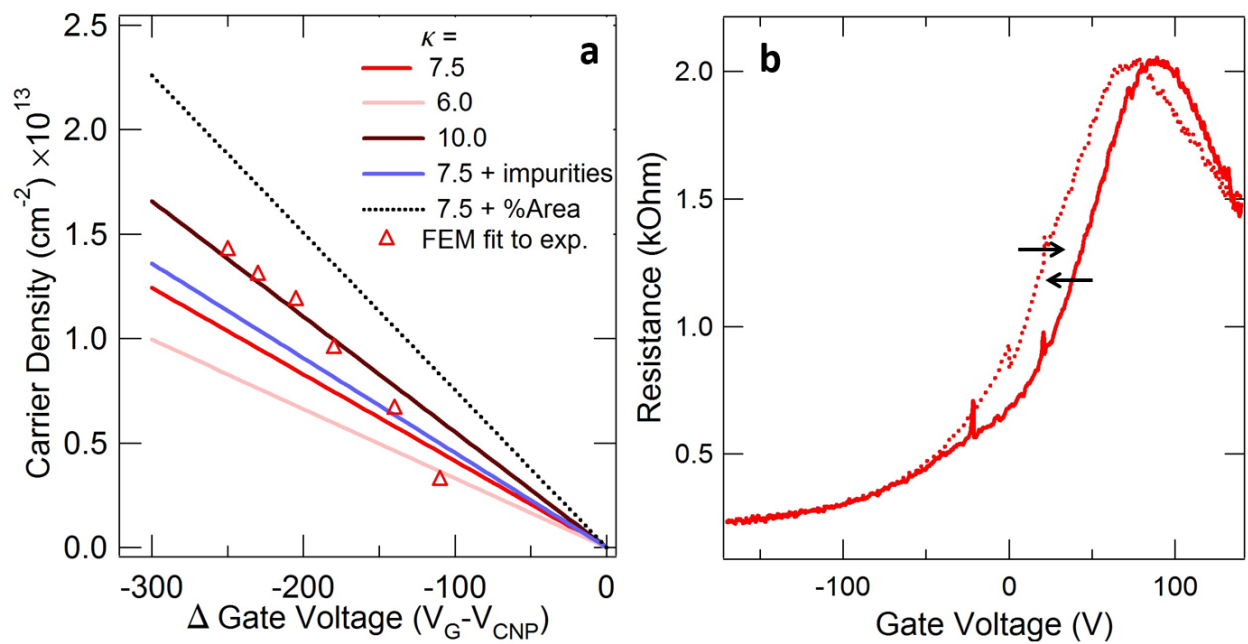

Figure S2: **Carrier density and resistance versus back gate voltage.** (a) Theoretical (lines) carrier density dependence on gate bias for a graphene FET device on a 1 ⎵m thick $SiN_x$ membrane with dielectric properties spanning those reported in literature.[15] The blue line indicates a graphene/$SiN_x$ device that includes estimated doping effects due to atmospheric and substrate impurities that have been reported in $SiO_2$.[7-10] The black dotted line models a graphene/$SiN_x$ device that contains a graphene surface patterned such that 45% of the sheet has been removed and the surface charge is concentrated into a smaller area. The triangles indicate the calculated carrier densities of our device determined by fitting the simulated peak position to the experimental results. (b) Hysteresis effects in graphene/SiN resistance as applied gate bias swept up (dotted line) and down (solid line). For (a, triangles), the CNP was assumed to occur at +80V, halfway between the two hysteric peaks.

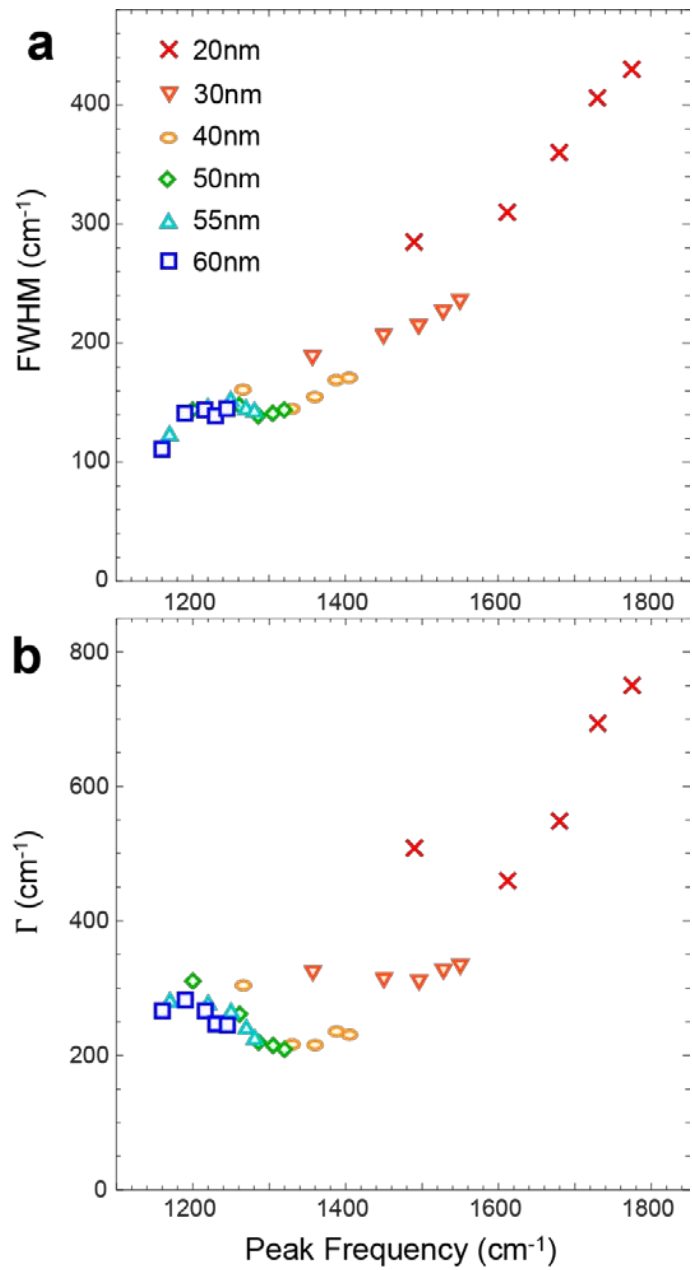

Figure S3: **Peak width and electron scattering rate.** Frequency dependence of (a) the full width at half maximum (FWHM) linewidth of the absorption peaks, and (b) the fitted electron scattering rate. The resonator width ranges from 20 to 60 nm and the carrier density varies from $0.66\times10^{13}$ to $1.42\times10^{13}$cm$^{-2}$.

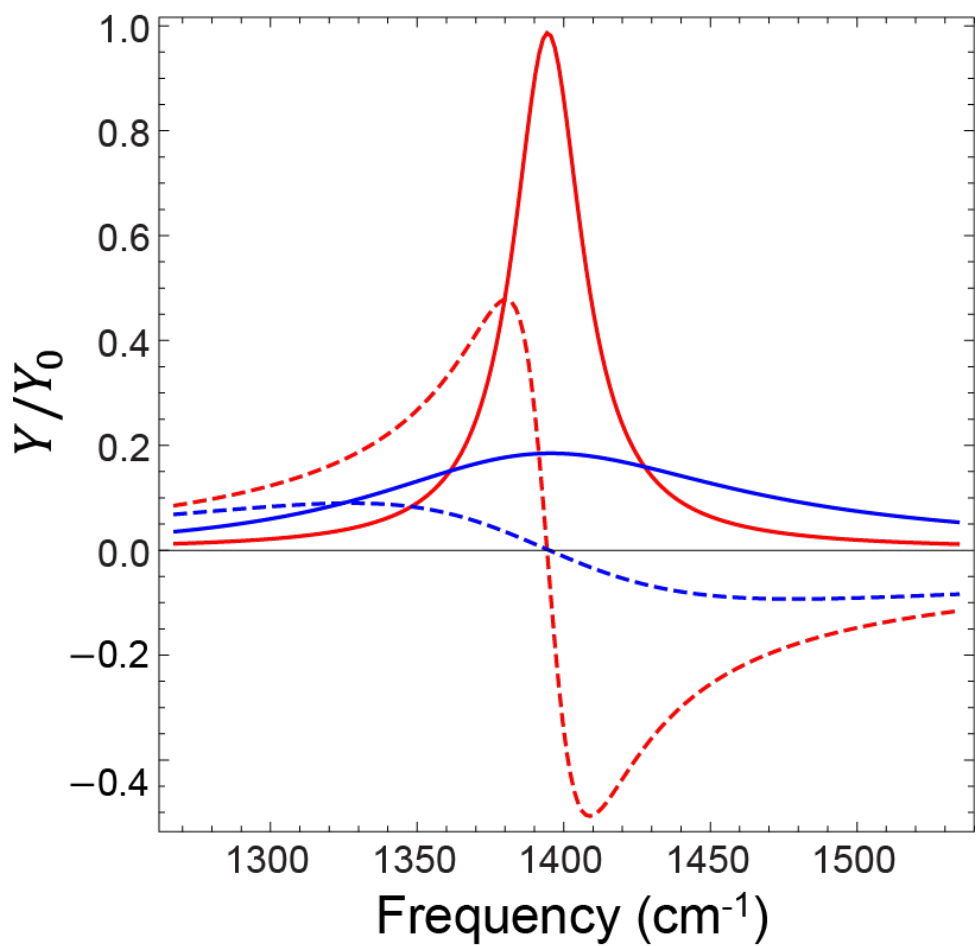

Figure S4: **Surface admittance versus frequency.** Frequency dependence of Re[$Y/Y_0$] (solid) and Im[$Y/Y_0$] (dashed) for $\mu = 550$ (blue) and 4,000cm$^2$/Vs (red). An array of infinitely long graphene nanoribbons (40nm width, 40nm spacing) is assumed. The carrier density is set to $1.42 \times 10^{13}$cm$^{-2}$.

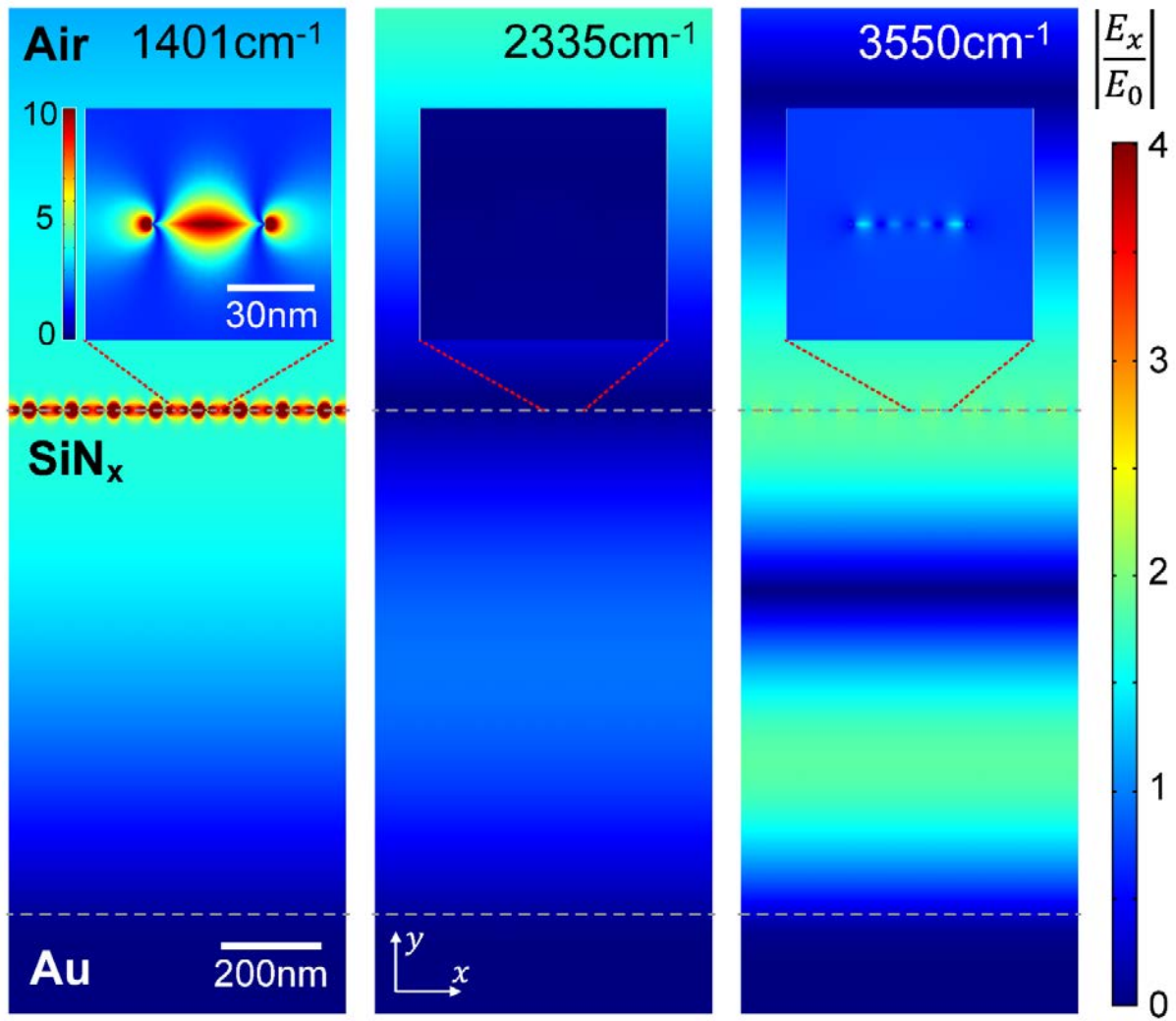

Figure S5: **Electric field distribution.** Theoretical electric field profile of a 40nm graphene nanoresonator with the highest achieved carrier density ($1.42\times10^{13}$cm$^{-2}$), obtained from an electromagnetic simulation assuming normal incidence. The quarter wavelength condition and plasmon resonance coincide at 1400cm$^{-1}$ (left). At 2335cm$^{-1}$, the optical thickness of SiN$_x$ is roughly half wavelength, resulting in vanishing electric field at the surface (middle). When the optical thickness of SiN$_x$ becomes three quarters of the wavelength, the surface electric field is maximized again, but the higher order plasmon resonance at this frequency is very weak (right).

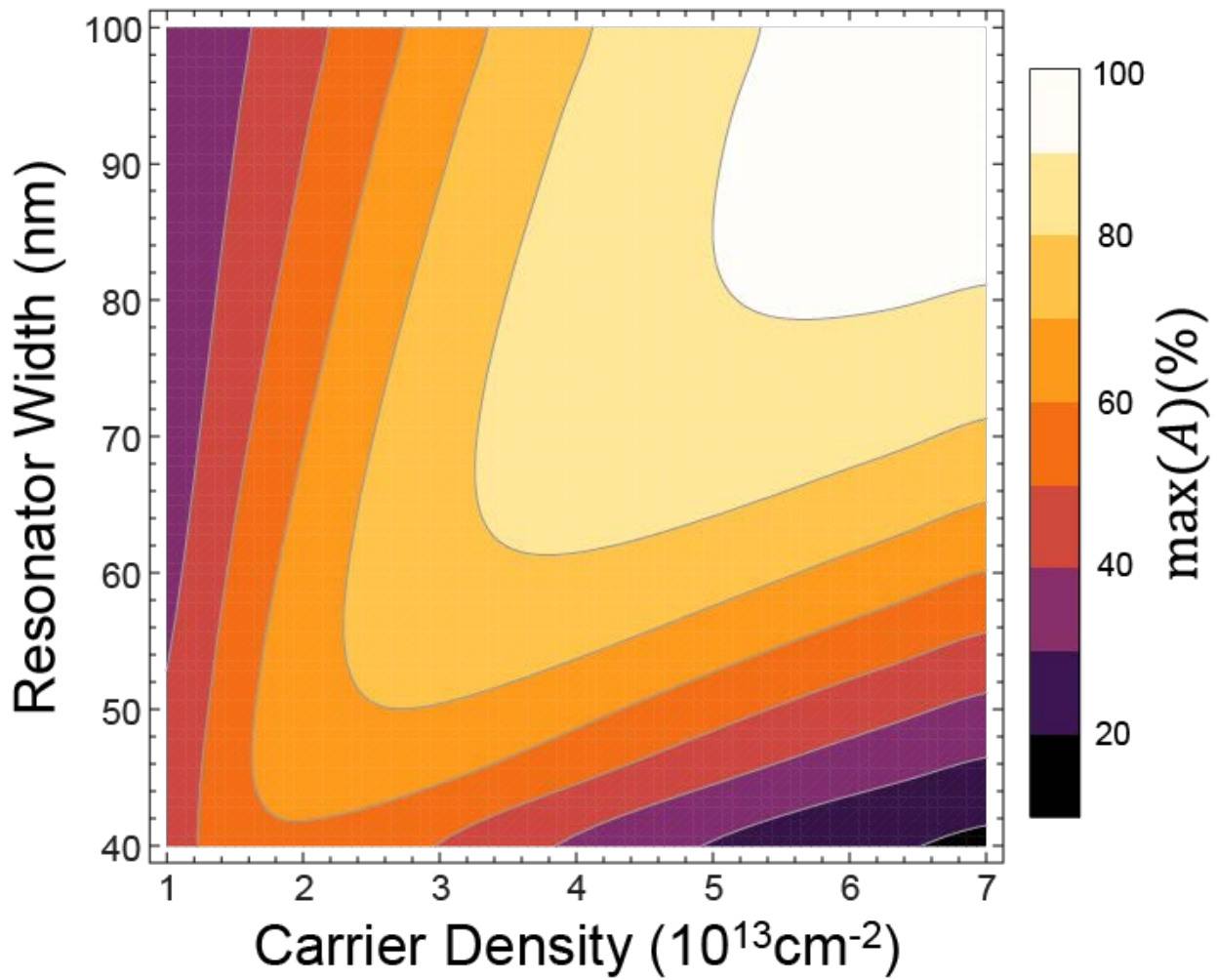

Figure S6: **Peak absorption versus carrier density and resonator width.** Maximum theoretical absorption in a graphene nanoribbon array as a function of the carrier density and the resonator width. The carrier density $1.0-7.0\times10^{13}$cm$^{-2}$, which is equivalent to the Fermi energy of 0.37−0.98eV. Increasing carrier density leads to better coupling between the incoming light and the graphene plasmons, resulting in stronger plasmon resonance. Therefore, higher doping tends to enhance the absorption performance. The spacing between ribbons is equal to the ribbon width and the SiN$_x$ thickness is set to 1um. The carrier mobility is assumed to be 550cm$^2$/Vs, and the interaction with graphene optical phonon is considered.[2-4]